\documentclass[prb,twocolumn,showpacs,preprintnumbers,amsmath,amssymb,superscriptaddress]{revtex4}

\usepackage{graphicx}
\usepackage{dcolumn}
\usepackage{bm}

\renewcommand*\vec[1]{\bm #1}

\begin{document}
\title{Relaxation properties of the quantum kinetics of
  carrier-LO-phonon interaction in quantum wells and quantum dots}

\author{P. Gartner}
\affiliation{Institute for Theoretical Physics,
             University of Bremen,
             28334 Bremen, Germany}
\affiliation{National Institute for Materials Physics, POB MG-7,
             Bucharest-Magurele, Romania}

\author{J. Seebeck}
\affiliation{Institute for Theoretical Physics,
             University of Bremen,
             28334 Bremen, Germany}

\author{F. Jahnke}
\email[Corresponding author: ]{www.itp.uni-bremen.de/~jahnke}
\affiliation{Institute for Theoretical Physics,
             University of Bremen,
             28334 Bremen, Germany}

\date{\today}

\begin{abstract}
  The time evolution of optically excited carriers in semiconductor
  quantum wells and quantum dots is analyzed for their interaction
  with LO-phonons. Both the full two-time Green's function formalism and
  the one-time approximation provided by the generalized
  Kadanoff-Baym ansatz are considered, in order to compare their
  description of relaxation processes.
  It is shown that the two-time quantum kinetics
  leads to thermalization in all the examined cases, which is not the
  case for the one-time approach in the intermediate-coupling regime,
  even though it
  provides convergence to a steady state. The thermalization criterion
  used is the Kubo-Martin-Schwinger condition.

\end{abstract}


\maketitle

\section{Introduction}
The Boltzmann equation is of central importance in many fields of
physics and, since its original formulation in the theory of gases it
has received a whole range of extensions to other domains
like plasma physics, nuclear physics, or semiconductors. In these
fields, Boltzmann scattering integrals are extensively used to model
relaxation and thermalization processes. Adapted versions of the
H-theorem ensure that, indeed, the equations describe the steady
evolution of the system towards the proper thermal equilibrium.
Early attempts to derive this irreversible behavior from the
quantum-mechanical evolution have shown \cite{vanHove:55} that the range of
validity of Boltzmann-like equations correspond to the low-coupling,
slowly-varying, long-time regime.

In more recent years with the experimental possibility to produce and
control transport and optical phenomena at ultra-short timescales,
quantum-kinetic theories\cite{Haug_Jauho:96,Schaefer_Wegener:02} have
been devised in order to describe rapid processes in which coherence
is still present, together with the
onset of dephasing and relaxation. This means that the kinetics has
to describe not only real-number quantities like
occupation probabilities, but also complex, off-diagonal
density-matrix elements, and their interference effects.

Not only fast dynamics, but also the necessity to extend the theory
beyond the weak-interaction limit has prompted the development of
quantum kinetics.
A typical example is provided by the interaction of carriers with
LO-phonons in semiconductor quantum dots, where a phonon bottleneck is
predicted by the Boltzmann result (see Ref.~\onlinecite{Benisty:95}
and references therein), in contrast to the quantum-kinetic treatment
of quantum-dot polarons in the strong-coupling regime\cite{Seebeck:05}
and many experimental findings.

The quantum-kinetic theory using non-equilibrium Green's functions (GF)
is one of the basic tools in this field. Its central object is the
one-particle, two-time GF, for which closed equations are
provided. Unfortunately, the large numerical effort needed for solving
these equations has limited previous applications of the two-time
formalism to the early-time regime.
The method has been used to describe the ultrafast optical excitation of
semiconductors where the interaction of carriers with
LO-phonons\cite{Hartmann:92}, the Coulomb
interaction of carriers\cite{Schaefer:96,Binder:97}, and their
combined influence\cite{Koehler:97} have been studied.
Calculations based on the two-time formalism also have been applied in
plasma physics\cite{Bonitz:96} and for nuclear
matter\cite{Koehler:01}.

Since the physically relevant information (e.g. population and
polarization dynamics) is contained in the one-time GF (the two-time
GF at equal times), it is clear that a closed equation for this
quantity would greatly simplify the procedure. This explains the huge
popularity of the generalized Kadanoff-Baym ansatz (GKBA) \cite{Lipavsky:86},
an approximation which expresses the two-time GF in terms of its
one-time component.

The GKBA has been extensively used in the past for a description of
non-Markovian contributions to ultra-fast relaxation and dephasing
processes.  Signatures of non-Markovian effects have been investigated
for the interaction of carriers with
LO-phonons\cite{Banyai:95,Banyai:96}. Furthermore, the built-up of
screening has been studied on the basis of a quantum-kinetic
description using the GKBA\cite{ElSayed:94,Banyai:98b} and included in
scattering calculations\cite{Vu:99,Vu:00}.
Results of the one- and two-time formulation have been compared for
early times addressing the carrier-carrier scattering\cite{Bonitz:96}
as well as the interaction of carriers with
LO-phonons\cite{Gartner:99}.

Boltzmann-like kinetic equations are obtained from the one-time theory
based on the GKBA by
further approximations: memory effects are neglected (Markov limit)
and free particle energies are used (low coupling limit). One encounters
therefore a situation in which only after taking two major
approximation steps, one reaches a kinetic theory for which the
physically expected correct relaxation behavior can be proven analytically.
To our knowledge, there is no attempt in the literature to
explore systematically the relaxation properties of either the
two-time formalism or its one-time approximation, despite their wide
applications and the obvious fundamental importance of the problem.
For example, the interest in laser devices based on quantum
wells\cite{Chow_Koch:99} and quantum dots \cite{QD} requires a good
understanding of the long-time behavior of the carriers in their
evolution to equilibrium.
Furthermore, the importance of non-Markovian effects in the
quantum-kinetic treatment of optical gain spectra for quantum-dot
lasers has been discussed recently.\cite{Schneider:04,Lorke:05}

In this paper, the relaxation properties in the long-time limit are
compared for the one-time and two-time quantum kinetics. As a test
case, we consider the interaction of carriers with LO-phonons in
semiconductor nanostructures, which is the dominant relaxation
mechanism for low carrier densities and elevated temperatures.
We study the optical excitation of quantum wells and quantum dots with
short laser pulses and calculate the dephasing of the coherent
polarization together with the relaxation and thermalization of the
excited carrier populations.
The equilibrium state of the interacting system is defined by the
Kubo-Martin-Schwinger condition.
We investigate if and under which conditions this equilibrium state is
reached in the time-dependent solution of the quantum kinetic models.
This provides a unique way to address the range of validity of the
involved approximations.

\section{Relaxation properties of the Boltzmann equation}

The Markovian limit of the kinetics, as described by the Boltzmann
equation, is a good example to start with, because its relaxation
properties are well understood and rigorously proven.
To be specific, we consider the Hamiltonian for the interacting system
of carriers and phonons,
\begin{align}
\label{eq:el-ph}
H_{\text{e-ph}} &=\sum_i \epsilon_i a^{\dagger}_i a_i
   + \sum_{\vec q} \hbar \omega_q b^{\dagger}_{\vec q} b_{\vec q} \nonumber\\
   &+ \sum_{i,j,\vec q} M_{i,j}(\vec q)a^{\dagger}_i a_j (b_{\vec q}+
    b^{\dagger}_{-\vec q}) ~,
\end{align}
where $i,j$ are indices for the carrier states and the momentum $\vec
q$ is the phononic quantum number.
The corresponding creation and annihilation operators for carriers and
phonons are given by $a^{\dagger}_i, a_i$ and $b^{\dagger}_{\vec q},
b_{\vec q}$, respectively.
The Boltzmann equation for the time evolution of the average
occupation number (population distribution) $f_i= \langle
a^{\dagger}_i a_i \rangle$ has the form
 \begin{equation}
\frac{\partial f_i}{\partial t} = \sum_j \left\{ W_{i,j} (1-f_i)f_j -
W_{j,i} (1-f_j)f_i \right\} \; ,
\label{eq:boltzmann}
\end{equation}
with the transition rates given by Fermi's golden rule
 \begin{align}
W_{i,j} &= \frac {2 \pi}{\hbar}\sum_{\vec q} |M_{i,j}(\vec q)|^2 \\
        &\times\left\{ N_{\vec q} \delta(\epsilon_i-\epsilon_j - \hbar
        \omega_{\vec q}) +
                      (N_{\vec q}\!+\!1)\delta(\epsilon_i-\epsilon_j +
        \hbar \omega_{\vec q}) \right\} \;.\nonumber
\end{align}
For a phonon bath in thermal equilibrium, $N_{\vec q}$ is a
Bose-Einstein distribution with the lattice temperature, and the
$\delta$-functions ensure the strict energy conservation in the $j
\rightarrow i $ transition process assisted by either the absorption
or the emission of a phonon.

The following properties of Eq.~(\ref{eq:boltzmann}) can be
analytically proven: (i) the total number of carriers $\sum_i f_i$ is
conserved, (ii) positivity is preserved, i.e., if at $t=0$ one has
$f_i \geqslant 0$ then this remains true at any later time, (iii) the
Fermi distribution $f_i= [e^{-\beta(\epsilon_i-\mu)} +1]^{-1} $ is a
steady-state solution of  Eq.~(\ref{eq:boltzmann}) and (iv) this
steady state is the large time limit of the solution $f_i(t)$ for any
positive initial condition {\it provided} a certain connectivity
property holds. This property is fulfilled if any state of the carrier
system can be reached from any other state through a chain of
transitions having non-zero rates. The temperature of the stationary
Fermi distribution is the lattice temperature, and the chemical
potential is fixed by the total number of carriers. If the set of
carrier states is not connected in the above sense, any connected
component behaves like a separate fluid and reaches equilibrium with
its own chemical potential.

As satisfying as this picture looks, several problems arise here. The
carrier-phonon interaction is essential as a relaxation mechanism but
the carrier energies themselves are taken as if unaffected by it. Both in
the energy conserving $\delta$-functions and in the final Fermi
distribution these energies appear as unperturbed. This corresponds to
a low-coupling regime, which may not be valid in practical
situations. Even in weakly polar semiconductors like GaAs, the
confined nature of the states in quantum wells (QWs) and even more so in
quantum dots (QDs), gives rise to an enhanced effective interaction
\cite{Inoshita:97}. For higher coupling constants one expects departures from
the simple picture discussed above. Moreover, in the case of a strong coupling
and with the inclusion of memory effects, neglected in the Markovian
limit, the energy conservation is not expected to hold.  Finally, and
specifically for LO phonons, their dispersionless spectrum, associated
with strict energy conservation turns the system into a disconnected
one. Indeed, each carrier can move only up and down a ladder with
fixed steps of size $\hbar \omega_{LO}$ but cannot jump on states
outside this ladder. A phonon bottleneck effect in QDs was predicted
on these grounds.\cite{Benisty:95}

\section{Statement of the problem}

It is clear that in most practical cases one has to turn to
quantum-kinetic treatments in which both energy renormalizations
and memory effects are considered. Such formalisms are provided by the
two-time Green's function kinetics or by one-time approximations
to it. In view of the discussion of the previous section, the
following questions, regarding the relaxation properties of the quantum
kinetics, are in order: (i) Is the particle number conserved? (ii) Is
positivity conserved? (iii) Is the system evolving to a steady state?
(iv) If yes, is this steady state a thermal equilibrium one?  In what
sense?

To our knowledge, with the exception of the first question, which can
be easily answered affirmatively, there is no definite and proven
answer available in the literature.
The aim of the present paper is to investigate how numerical solutions
of the quantum-kinetic equations for realistic situations behave in
the discussed respects.
For this purpose, we compare the results of the two-time and the one-time approach.

\section{Two-time quantum kinetics}

In this section we specify the Hamiltonian, Eq.~(\ref{eq:el-ph}), for
the case of a homogeneous two-band semiconductor, where carriers
interact with LO-phonons via the Fr\"ohlich coupling,
\begin{align}
 H_{\text{e-ph}} &=\sum_{\vec k,\lambda}\epsilon^{\lambda}_{\vec
 k}\;a^{\dagger}_{\vec k,\lambda}a_{\vec
 k,\lambda} + \sum_{\vec q} \hbar \omega_q\; b^{\dagger}_{\vec q} b_{\vec q} \nonumber\\
 &+ \sum_{\vec k,\vec q, \lambda} g_{q}\; a^{\dagger}_{\vec
           k+\vec q,\lambda}a_{\vec k,\lambda}(b_{\vec
 q}+b^{\dagger}_{-\vec q}) ~.
\end{align}
The carrier quantum numbers are the band index $\lambda=c,v$ and the
3D- (for the bulk case) or 2D- (for QWs) momentum $\vec k$.
The coupling is defined by $g_{q}^2 \sim \alpha / q^2$  for the 3D
case, or by $g_{q}^2 \sim \alpha F(q)/ q$ for the quasi-2D case, with
the form factor $F(q)$ related to the QW confinement function and the
Fr\"ohlich coupling constant $\alpha$.
Additional terms to this Hamiltonian describe the optical excitation
and the Coulomb interaction in the usual
way.\cite{Schaefer_Wegener:02}
We consider only sufficiently low excitations so that carrier-carrier
scattering and screening effects are negligible.
Then the only contribution of the Coulomb interaction is the
Hartree-Fock renormalization of the single particle energies and of
the Rabi frequency.

The object of the kinetic equations is the two-time GF,
$G_{\vec k}^{\lambda, \lambda'}(t_1,t_2)=G_{\vec k}^{\lambda,
  \lambda'}(t,t-\tau)$. We use the parametrization of the two-time
plane $(t_1,t_2)$ in terms of the main time $t$ and relative time
$\tau$. One can combine the two Kadanoff-Baym equations \cite{Danielewicz:84}
which give the derivatives of the GF with respect to $t_1$ and $t_2$,
according to $\partial / \partial t=\partial / \partial t_1+\partial /
\partial t_2 $ and $\partial / \partial \tau=-\partial / \partial t_2
$ in order to  propagate the solution either along the time diagonal
($t$-equation) or away from it ($\tau$-equation). As two independent
GFs we choose the lesser and the retarded ones, and limit ourselves to the
subdiagonal halfplane $\tau \geqslant 0$, since supradiagonal
quantities can be related to subdiagonal ones by complex
conjugation. With these options and in matrix notation with respect to
band indices, the main-time equation reads
\begin{align}
\label{eq:teq}
  i\hbar\frac{\partial}{\partial t}G^{R,<}_{\vec k}(t,t-\tau)
  &=\Sigma^{\delta}_{\vec k}(t) ~ G^{R,<}_{\vec k}(t,t-\tau) \nonumber\\
  &-G^{R,<}_{\vec k}(t,t-\tau) ~ \Sigma^{\delta}_{\vec k}(t-\tau) \nonumber\\
  &+\left.i\hbar\frac{\partial}{\partial t}G^{R,<}_{\vec
  k}(t,t-\tau)\right |_{\text{coll}} \; ,
\end{align}
where the instantaneous self-energy contains the external and the
self-consistent field,
\begin{equation}
  \Sigma^{\delta}_{\vec k}(t) = \left(
    \begin{array}{cc}
     \epsilon^c_{\vec k} & -\hbar \Omega_R(t)
     \\ -\hbar \Omega^*_R(t) & \epsilon^v_{\vec k}
    \end{array}
    \right)
    + i \hbar \sum_{\vec q} V_q \; G^<_{\vec k-\vec q}(t,t) \; .
\end{equation}
The collision term in Eq.~(\ref{eq:teq}) has different expressions for
$G^R$ and $G^<$,
\begin{align}
\label{eq:coll_ret}
\left.i\hbar\frac{\partial}{\partial t}G^{R}_{\vec
 k}(t,t-\tau)\right|_{\text{coll}} &=
 \int_{t-\tau}^t dt' \Big[ \Sigma^R_{\vec k}(t,t')G^R_{\vec
 k}(t',t-\tau)\nonumber \\
 &-G^R_{\vec k}(t,t')\Sigma^R_{\vec k}(t',t-\tau)\Big] ,
\end{align}
\begin{align}
\label{eq:coll_less}
\left.i\hbar\frac{\partial}{\partial t}G^{<}_{\vec
   k}(t,t-\tau)\right|_{\text{coll}} &= \int^{t}_{-\infty}d t'\Big[
   \Sigma_{\vec k}^R G_{\vec k}^<
   +\Sigma_{\vec k}^< G_{\vec k}^A \nonumber\\
   &- G_{\vec k}^R \Sigma_{\vec k}^<
   - G_{\vec k}^< \Sigma_{\vec k}^A
   \Big] .
\end{align}
The time arguments of the self-energies and GFs in
Eq.~(\ref{eq:coll_less}) are the same as in Eq.~(\ref{eq:coll_ret})
and are omitted for simplicity.
The advanced quantities are expressible through retarded ones by conjugation.
The self-energies are computed in the self-consistent RPA scheme and
have the explicit expressions
\begin{align}
\Sigma^R_{\vec k}(t,t') &= i \hbar \sum_{\vec q} g^2_{q} \; \big[
                        D_{\vec q}^>(t-t')G^R_{\vec k -\vec q}(t,t')
                        \nonumber\\
                        &+ D_{\vec q}^R(t-t') G^<_{\vec k -\vec
                        q}(t,t')\big] ~,\nonumber \\
                        \nonumber\\
\Sigma^<_{\vec k}(t,t') &=i \hbar \sum_{\vec q} g^2_{q}\; D_{\vec
                        q}^<(t-t')G^<_{\vec k -\vec q}(t,t') ~,
\end{align}
with the equilibrium phonon propagator
\begin{align}
   D^{\gtrless}_{\vec q}(t) &=-\frac{i}{\hbar} ~\Big[ ~N_{\vec q}
                     ~e^{\pm i\hbar\omega_{\vec q} t}
                     +~ (1+N_{\vec q}) ~e^{\mp i\hbar\omega_{\vec q}
                     t} ~\Big] ~.
\end{align}
For practical calculations, we use dispersionless phonons
$\omega_{\vec q}=\omega_{LO}$.

\begin{figure}[htb!]
   \includegraphics[angle=0, width=0.8\columnwidth]{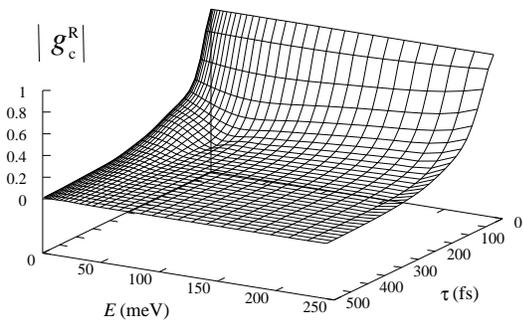}
   \caption{Absolute value of the polaronic retarded GF for electrons
     in a CdTe QW at T = 300K. $E=0$ corresponds to the conduction
     band edge.}
    \label{pol_gf}
\end{figure}

The above set of equations has to be supplemented by specifying the
initial conditions.
For all the times prior to the arrival of the optical pulse, the
system consists of the electron-hole vacuum in the presence of the
phonon bath.
This is an equilibrium situation, characterized by diagonal GFs, which
depend only on the relative time.
More precisely, one has
\begin{align}
\label{eq:vacGF}
  G^R(t,t-\tau) &= \left(
    \begin{array}{cc}
     g^R_c(\tau) & 0
     \\ 0 & g^R_v(\tau)
    \end{array}
    \right) ~,
   \nonumber\\\nonumber\\
  G^<(t,t-\tau) &= \left(
    \begin{array}{cc}
     0 & 0
     \\ 0 & - g^R_v(\tau)
    \end{array}
   \right) ~,
\end{align}
where $g^R$ is the retarded GF of the unexcited system. The calculation of
this GF is an equilibrium problem, namely that of the Fr\"ohlich
{\it polaron}. The polaronic GF is the solution of the $\tau$-equation,
\begin{equation}
\left(i\hbar\frac{\partial}{\partial \tau}-\epsilon_{\vec{k}}^\lambda
\right)g^{R}_{\vec{k}, \lambda}(\tau)=\int_0^\tau d\tau'\;
      \Sigma^{R}_{\vec{k}, \lambda}(\tau-\tau')g^{R}_{\vec{k},
	\lambda}(\tau') \; ,
\end{equation}
in which, to be consistent with the $t$-evolution described above, the
RPA self-energy is again used. The vacuum GF of Eq.~(\ref{eq:vacGF})
is not only the starting value for the GF in the main-time evolution
but it also appears in the integrals over the past which require GF
values before the arrival of the optical pulse. Moreover, the
presence of the polaronic GF brings into the picture the complexities
of the spectral features of the polaron, with energy renormalization
and phonon satellites. Finally, the decay of the polaronic GF
introduces a natural memory depth into the problem. An example is seen in
Fig.~\ref{pol_gf} where, due to a rather strong coupling constant and
a high temperature, the decay with the relative time $\tau$ is rapid.
This allows to cut the infinite time integrals of
Eq.~(\ref{eq:coll_less}) at a certain distance away from the diagonal.

\begin{figure}[htb!]
   \includegraphics[angle=0, width=0.8\columnwidth]{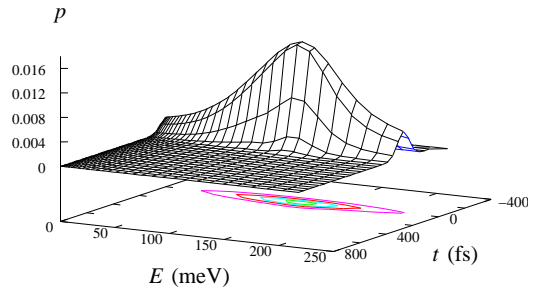}
   \caption{(Color online) Time evolution of the coherent interband polarization
   after optical excitation of a CdTe QW with a 100 fs laser pulse
   centered at time $t=0$, using a two-time calculation.}
   \label{pol}
\end{figure}

In Fig.~\ref{pol_gf} the momentum argument is replaced by
the unrenormalized electron energy $E=\hbar^2 k^2/2 m^*_e$. This
change of variable is allowed by the fact that the momentum dependence
of the polaronic GF is isotropic. The same energy argument is used in
the subsequent figures, with the exception of Fig.~\ref{pol} where the
reduced mass is employed ($E=\hbar^2 k^2/2 m^*_r$) as being more
appropriate in the case of the polarisation. The choice of an
energy variable facilitates the comparison with other energies
appearing in the theory. For instance, in Fig.~\ref{pol_gf} a somewhat
slower decay is seen at low energies and is a trace of the phonon
threshold ($\hbar \omega_{LO}=$ 21 meV). Also, in
Figs.~\ref{pol},\ref{fe} below, the functions have a peak around the
detuning (120meV).

With the specification of the initial conditions, the problem to be
solved is completely defined. After obtaining the two-time GFs, the
physically relevant information is found in the equal-time lesser GF,
which contains the carrier populations and the polarization.
This program is carried out for the case of a CdTe ($\alpha=0.31$) QW at room
temperature. The excitation conditions are defined
by a Gaussian-shaped pulse of 100 fs duration (FWHM of the
intensity), having an excess energy of 120meV above the unrenormalized
band gap and an area of 0.05 (in $\pi$ units). This gives rise to
carrier densities in the order of $10^9/\mathrm{cm}^2$, sufficiently
low to neglect carrier-carrier scattering.

\begin{figure}[htb!]
   \includegraphics[angle=0, width=0.8\columnwidth]{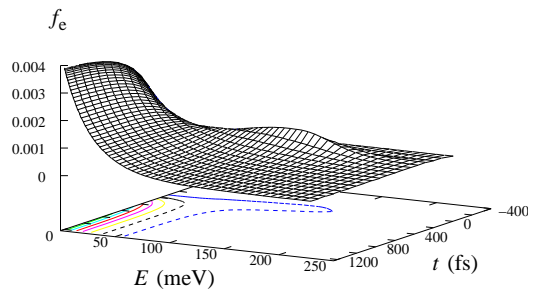}
   \caption{(Color online) Time evolution of the electron population distribution for
   the same situation as in Fig.~\ref{pol}.}
   \label{fe}
\end{figure}

As seen in Figs.~\ref{pol} and \ref{fe}, the interaction of carriers
with LO-phonons provides an efficient dephasing and leads, in a
sub-picosecond time
interval, to a relaxation of the electron population into a
steady-state distribution. The same is true for the hole population (not
shown). Before discussing this result we compare it to the outcome of
the one-time calculation.

\section{One-time quantum kinetics}

To obtain from the Kadanoff-Baym equations for the two-time GF  a
closed set of equations for the
equal-time lesser GF, $G^<(t,t)$, one can use  the generalized
Kadanoff-Baym ansatz (GKBA)\cite{Lipavsky:86}. The ansatz reduces the
time-offdiagonal GFs
appearing in the collision terms of Eq.~(\ref{eq:coll_less})
to diagonal ones with the help of the spectral (retarded or advanced)
GFs. Therefore, the GKBA has to be supplemented with a choice of
spectral GFs. In our case, it is natural to use for this purpose the
polaronic GF. For $\tau\geq 0$, this leads to the GKBA in the form
\begin{eqnarray}
G^<(t,t-\tau) \approx  i \hbar \;g^R(\tau) \; G^<(t-\tau,t-\tau) \; .
\end{eqnarray}

\begin{figure}[htb!]
   \includegraphics[angle=0, width=0.8\columnwidth]{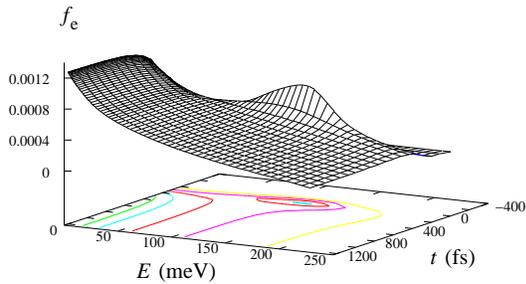}
   \caption{(Color online) Time evolution of the electron population distributions in
   a CdTe QW, using the same excitation conditions as in
   Figs.~\ref{pol},\ref{fe} and a one-time calculation.}
   \label{fe1t}
\end{figure}

The result of this procedure for the same system and using the same
excitation conditions as for the two-time calculation is shown in
Fig.~\ref{fe1t}. We find that the steady state obtained in this
way differs appreciably from that of the two-time calculation.

\section{The KMS condition}

For a fermionic system in thermal equilibrium, the following
relationship connects the lesser and the spectral GF \cite{Haug_Jauho:96}
\begin{align}
G^<_{\vec k}(\omega) &= -2i ~f(\omega) ~\text{Im} ~G^R_{\vec
  k}(\omega) ~,\nonumber\\
f(\omega) &= \frac{1}{e^{\beta(\hbar \omega - \mu)}+1} ~.
\label{eq:kms}
\end{align}
In thermodynamic equilibrium, the GFs depend only on the relative time and
Eq.~(\ref{eq:kms}) involves their Fourier transform with respect to
this time. The relationship is known as the Kubo-Martin-Schwinger
(KMS) condition or as the fluctuation-dissipation theorem, and leads
to a thermal equilibrium population given by
\begin{eqnarray}
f^{\lambda}_{\vec k} = - \int \frac{d \hbar
  \omega}{\pi}f(\omega)~\text{Im}~G^{R,\lambda \lambda}_{\vec
  k}(\omega)\;  .
\label{eq:kmspop}
\end{eqnarray}
The two-time theory provides the excitation-dependent retarded GF
along with the lesser one, the formalism being a system of coupled
equation for these two quantities.
Nevetheless, in the low excitation regime used here the difference
between the actual retarded GF and its vacuum counterpart $g^R_{\vec
  k, \lambda}(\omega)$ turns out to be negligible, as can be checked
numerically. Therefore, the latter can be used in
Eq.~(\ref{eq:kmspop}) without loss of accuracy.

\begin{figure}[htb!]
   \includegraphics[angle=0, width=0.8\columnwidth]{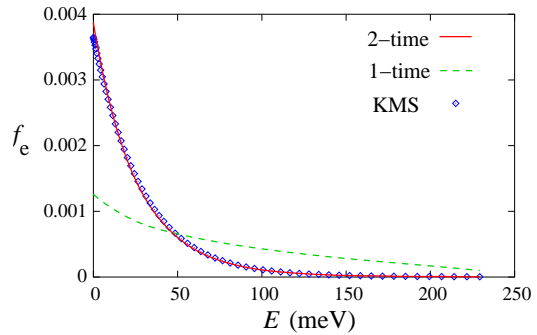}
   \caption{(Color online) One- and two-time CdTe QW electron populations at t = 1240 fs and
  the KMS result.}
   \label{kms_a}
\end{figure}

The thermal equilibrium distribution $f^{\lambda}_{\vec k}$ obtained
from the KMS condition is the generalization of the Fermi function of
the non-interacting case and is used as a check of a proper
thermalization.
The test of the two steady-state solutions against the KMS
distribution function is seen in Fig.~\ref{kms_a}.
The two-time calculation is in
good agreement with the KMS curve, but the one-time evolution is
not. It appears that the one-time kinetics produces a steady state
with a temperature considerably exceeding that of the phonon bath.

\begin{figure}[htb!]
   \includegraphics[angle=0, width=0.8\columnwidth]{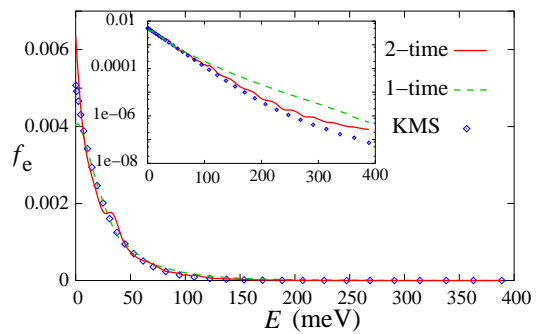}
   \caption{(Color online) Electron population at $t$=1600 fs for a GaAs QW and
            optical excitation with a 100 fs laser pulse at $t=0$.
            Solutions of the two-time and the one-time quantum
            kinetics are compared with the KMS result.
            Inset: same on semi-logarithmic scale.}
   \label{kms_c}
\end{figure}

It is to be expected, however, that for a weaker coupling the
discrepancy between the full two-time procedure and the GKBA is less
severe. This is indeed the case, as shown in Fig.~\ref{kms_c},
where results for a GaAs ($\alpha = 0.069$) QW are given. The wiggles
seen in the two-time curve are traces of the phonon cascade, which are still
present. This is due to the much longer relaxation time in
low-coupling materials. Nevertheless the trend is clear, the
steady-state solutions of both
approaches are in good agreement with the KMS condition.

\begin{figure}[htb!]
   \includegraphics[angle=0, width=0.8\columnwidth]{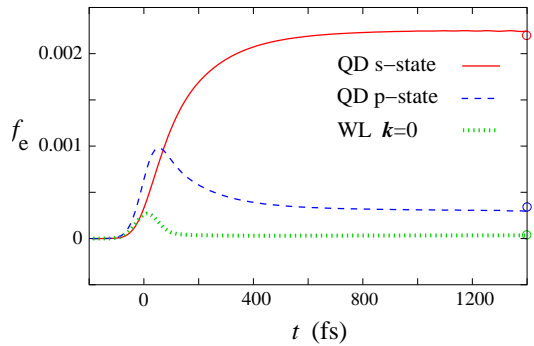}
   \caption{(Color online) Electron populations in the localized $s$ and $p$ states of a
  CdTe QD and in the extended $\vec{k}=0$ WL state after optical
  excitation with a 100 fs laser pulse at $t=0$, as calculated using
  the two-time kinetics. Open circles represent the equilibrium values
  according to the KMS condition.}
   \label{qd2t}
\end{figure}

\begin{figure}[htb!]
   \includegraphics[angle=0, width=0.85\columnwidth]{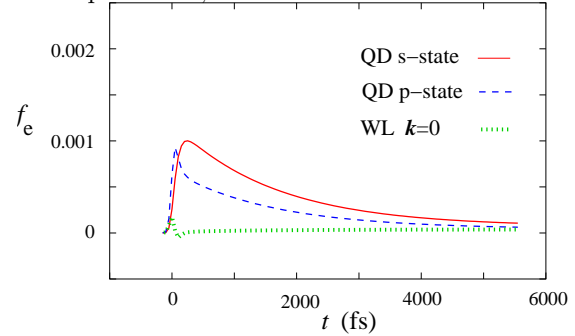}
   \caption{(Color online) Same as Fig.~\ref{qd2t}, but using the one-time kinetics.
   Note that an identical ordinate axis is used to facilitate the comparison.}
   \label{qd1t}
\end{figure}

Another important example concerns a non-homogeneous system. It
consists of CdTe lens-shaped self-assembled QDs, having both for
electrons and for holes
two discrete levels below the wetting-layer (WL) continuum. These
states are labelled $s$ and $p$, according to their $z$-projection
angular momentum.  We consider an equidistant energy spacing of  $2.4 \hbar
\omega_{LO}$ between the WL continuum edge, the $p$-level and the
$s$-level, for the electrons and a  $0.27 \hbar \omega_{LO}$ similar
spacing for holes. The formalism used is the same as for the
homogeneous systems but with the momentum replaced by a state
quantum number running over the discrete QD states and the WL
continuum. This amounts to neglecting GF matrix elements which are
off-diagonal in the state index, but still keeping off-diagonal terms
with respect to the band index. This has been shown to be a reasonable
approximation for QDs. \cite{Seebeck:05,Kral:98}
Our calculations for this example include both localized QD and
delocalized WL states.
We consider a harmonic in-plane confinement potential for the
localized states and construct orthogonalized plane waves for the
delocalized states in the WL plane.
The strong confinement in growth direction is described by a step-like
finite-height potential.
Details for the calculation of interaction matrix elements are given
in Ref.~\onlinecite{Seebeck:05}.
In Figs.~\ref{qd2t} and \ref{qd1t}, the time evolution of the
population of electrons is shown. The system is pumped close to the
renormalized $p$-shell energy with a 100 fs laser pulse at time
$t=0$. Therefore the majority
of the carriers is initially found in the $p$-state (which has a
two-fold degeneracy due to the angular momentum in addition to the
spin degeneracy). Nevertheless,
efficient carrier relaxation takes place, even if the level spacing
does not match the LO-phonon energy, and a steady state is reached. The
two-time results are again in agreement with the KMS condition, shown
by open circles. The one-time evolution shows a non-physical intermediate
negative value for the WL population and converges to a state in
strong disagreement with the KMS result.

\section{Conclusions}
The long-time behaviour of different quantum kinetic approaches to the
problem of carrier-scattering by means of LO-phonons was analyzed, in order to
assess their relaxation properties. As a test of proper convergence to
thermal equilibrium, the KMS condition was used. We considered
materials with low (GaAs) and intermediate (CdTe) Fr\"ohlich
coupling. The results can be summarized as follows: (i) In both the
one-time and the two-time quantum kinetics steady states are reached.
(ii) The steady state produced by the two-time approach obeys
the KMS condition in all cases considered. (iii) The one-time result
agrees with the KMS condition only at low coupling and differs
considerably for larger ones.

\section*{Acknowledgments}
This work has been supported by the Deutsche Forschungsgemeinschaft (DFG).
A grant for CPU time at the  Forschungszentrum J\"ulich is gratefully
acknowledged.


\end{document}